\documentstyle[preprint,eqsecnum,aps]{revtex}
\newcommand{\beq}{\begin{equation}}
\newcommand{\eeq}{\end{equation}}
\newcommand{\beqa}{\begin{eqnarray}}
\newcommand{\eeqa}{\end{eqnarray}}

\newcommand{\kBT}{\mbox{$k_{\rm B}T$}}

\begin{document}
\draft
\title{Statistical Mechanics of the Vicinal Surfaces with Adsorption
}
\author{Noriko Akutsu$^1$, Yasuhiro Akutsu$^2$ and Takao Yamamoto$^3$\\
}
\address{
$^1$Faculty of Engineering, Osaka Electro-Communication
 University,
Hatsu-cho, Neyagawa, Osaka 572-8530, Japan\\
$^2$Department of Physics, Graduate School of Science, Osaka University,
Machikaneyama-cho, Toyonaka, Osaka 560-0011, Japan\\
$^3$Department of Physics, Faculty of Engineering, Gunma University,
Kiryu, Gunma 376-0052, Japan\\
}
\date{\qquad \qquad \qquad 
}
\maketitle
\begin{abstract}
We study the vicinal surface with adsorption below the roughening temperature, using the restricted solid-on-solid model coupled with the Ising model.  By the product-wavefunction renormalization group method, we calculate   the surface gradient $p$ and the adsorption coverage $\Theta$ as a function of the  Andreev field $\eta$ which makes surface tilt.
Combining Monte Carlo calculations, we show that  there emerges effective attraction between the steps.
This attractive interaction leads to instability against step bunching. \\


\end{abstract}
%
%

%

\narrowtext
\newpage

\section{Introduction}

Below the roughening temperature $T_{\rm R}$, the vicinal surface which is a slightly tilted surface relative to one of the facet plane of a crystal, is well described in terms of terraces, steps and kinks (TSK picture).  Under the picture, the vicinal surface is known to show Gruber-Mullins-Pokrovsky-Talapov (GMPT) universal behavior\cite{gmpt}.
The surface free energy (per projected area), then, is written as~\cite{gmpt,haldane,izuyama,jayaprakash,schults,pimpinelli}   
\begin{equation}
f(\rho)=f(0)+\gamma \rho +B \rho^3 +O(\rho^4). \label{GMPTform}
\end{equation}
where $\rho$ is the step density, $\gamma$ is the step tension, and $B$ represents the effect of step-step interactions.  The step density $\rho$ relates to the surface gradient $p$ (along a crystal axis) as $p = \tan \phi = \rho a_d$ ($\phi$ is the tilt angle of the vicinal surface, and $a_d$ is the height of a single step).  For convenience we take the units where $a_d=1$ in the present article.  To be precise, $\gamma$ and $B$ depend on the mean running direction angle (relative to one of the crystal axes on the facet plane) of the steps, which we  denote by $\theta$; we should then write $\gamma =\gamma(\theta,T)$ and $B =B(\theta,T)$ ($T$: temperature).

From the thermodynamics of equilibrium crystal shape, the GMPT form of the vicinal surface free energy leads to
\beq
\eta=\frac{\partial f(p)}{\partial p }=\gamma  +3 B p^2 +O(p^3), \label{eta}
\eeq
where $\eta$ is Andreev field\cite{andreev} which makes the surface tilt.

It has been known that adsorbed atoms often change properties of the surface~\cite{pimpinelli,eaglesham,desjonqueres,copel,jones,fujita,williams98,hannon,latyshev,ozcomert,vonhoegen98}.
To discuss the adsorption effect on the vicinal surface, we take the restricted solid-on-solid (RSOS) model\cite{RSOS} on the square lattice, coupled with the Ising spin system representing the adsorbed gas\cite{aay99}.  
The effect of vapor pressure of adsorption in environmental gas phase is taken into consideration through the surface chemical potential of adsorbates.

\section{THE RSOS-ISING COUPLED MODEL}

In the RSOS model, we restrict each nearest-neighbor (nn) height difference $\Delta h$ to be $\Delta h=0,\pm1$, which is a reasonable simplification because configurations with large $|\Delta h|$ are energetically unfavorable.  

We assume that the gas atoms are likely to adsorb at step edge positions and that the adsorbed atom modifies the ledge energy locally.   Ferromagnetic interactions (attractive interaction, in the lattice-gas picture) with coupling constant $J$ between the nearest-neighbor spins on the rotated square lattice is considered.  We assume simple linear modification of the ledge energy as $\epsilon\rightarrow\epsilon (1-\alpha \sigma)$ ($\sigma$: Ising spin).  This modification leads to the interaction between the RSOS model and the Ising model.

The Hamiltonian of the RSOS-Ising coupled model is, therefore, written as
\beqa
{\cal H}=\sum_{<i,j>} \epsilon (1-\alpha \sigma_{b(i,j)}) 
|h_i-h_j|   \nonumber \\
- J \sum_{<b,b'>} \sigma_{b} \sigma_{b'}
-H \sum_{<b,b'>} \sigma_{b},
\label{hamil}
\eeqa
where $h_i$ is the integer surface height at site $i$, $\epsilon$ 
the ``bare'' ledge energy, $\sigma_{b(i,j)} = \pm 1$ the Ising 
spin variable on the bond $b(i,j)$ connecting the nn site pair 
$<i,j>$, and $H$ is an external field corresponding to the surface chemical potential for adsorbate.  We should note that the RSOS condition ($|\Delta h| 
\leq 1$ for each nn site pair) is implicit in (\ref{hamil}). (Fig. 1)

We set Cartesian coordinate  system to the square lattice so that the mean running direction of steps becomes $y$ direction.
The Andreev field $\eta$\cite{andreev} along the $x$-direction to tilt the surface is introduced by adding a term $-\eta\sum_{m,n}(h_{(m+1,n)}-h_{(m,n)})$ in the Hamiltonian (\ref{hamil}) ($(m,n)$ is the position vector of the lattice site).


The Andreev surface free energy  $\tilde{f}(\eta)$ which  directly gives the equilibrium crystal shape (ECS)\cite{andreev} is  calculated from the model Hamiltonian (\ref{hamil}) with the Andreev-field term by
\beq
\tilde{f}(\eta)= - \frac{1}{N} \kBT \ln Z, \label{tildef}
\eeq
where $Z$ is the partition function of the RSOS-Ising system.

We calculated the partition function  by the transfer-matrix method.  
We extend the  mapping between the RSOS model and the vertex model on the dual lattice,\cite{RSOS,vanBeij} to obtain a ``decorated'' vertex model (Fig.1) which is regarded as a  $19\times16=304$ vertex model.
For approximate diagonalization of the transfer matrix, we employ the 
product-wave-function renormalization group algorithm (PWFRG)\cite{pwfrg,HOA} 
which is a variant of the White's density-matrix 
renormalization group (DMRG) method\cite{white}.

In the vertex-model representation, the surface gradient $p$ along the $x$-direction is just the thermal average of the vertical edge variable of the vertex model, which can easily be calculated from the fixed-point wavefunction obtained by the PWFRG.
By sweeping the field $\eta$, we obtain a $p-\eta$ curve  which plays the role of  the ``equation of state''. Actual calculation is very similar to that of the magnetization curve for spin chains\cite{HOA,OHA} (see also \cite{honda-horiguchi}).

The Ising magnetization $M$ which is calculated by $<\sigma_b>$ where $< \cdot >$ denotes the thermal average.
The adsorption coverage $\Theta$ is connected to $M$ by $\Theta = (M+1)/2$.

\section{ADSORPTION MEDIATED INTER-STEP ATTRACTION}
In the RSOS-Ising coupled system, we find, at low temperature, the crystal shape shows the ``first-order" transition\cite{rottman84} at the facet edge, that is, the surface gradient $p$ jumps from 0 to a finite value $p_c(T, H)$ at the facet edge.
In Fig. 2, we show  $p-\eta$ curve and $M-\eta$ curve calculated by PWFRG method at $J/\epsilon=0.15$, $H/\epsilon = -0.01$.
Both $M$ and $p$ jump at facet edge at low temperatures.

We also have made Monte Carlo calculation at  $J/\epsilon=0.15$, $H/\epsilon = -0.01$ and temperature $\kBT/\epsilon = 0.4$.
In  Fig. 3, we show the top view of RSOS surface (above)  and the adsorbates configuration (below).  Steps get together to form a ``quasi-facet'', and the surface gradient of the quasi-facet agrees with the $p_c(0.4, -0.01)$ in Fig. 2.
The vicinal surface of a mean surface gradient $p$ breaks up into two regions: one with  $p=0$ and the other with $p=p_c$.  This phenomenon is precisely the phase separation characteristics to the first-order transition~\cite{williams91}.
We should remark that the quasi-facet with $p= p_c$ is not a ``true''  facet in the ECS drawn by the Wulff construction. 
The facet must be the singular surface corresponding to the cusp orientation of the Wulff figure\cite{wulff,rottman84}; this is not the case for the quasi-facet.
Although the quasi-facet is rough, the boundary between the  $p=0$ surface and the quasi-facet (``quasi-facet edge'') is ``sharp''.

From a microscopic point of view, the above thermal step bunching implies the existence of attractive interaction between the steps.  
To confirm this, we calculate a  ``terrace width distribution function'' $P(l)$ of the RSOS-Ising system (number of steps $= 2$) by Monte Carlo method up to $4 \times 10^7$MCS (Fig. 4).
We denote locations of step ledge at $y$ of the step 1 and the step 2 by $x_1(y)$ and $x_2(y)$, respectively. 
The linear size of the square lattice in the $x$ direction and the $y$ direction is denoted by  $L_x $ and $L_y $, respectively.
We impose the periodic boundary conditions along both $x$- and $y$-directions. To be precise, the ``periodic boundary condition'' along the $x$-direction is the one where the uppermost terrace is connected to the lowermost terrace. The ``terrace width'' $l$ is calculated to be  $|x_2(y) - x_1(y)|$ or $|x_1(y) + L_x - x_2|$.

The solid curve in Fig. 4 clearly shows double peaks with very narrow peak  width (this shape is unchanged under the increase of  $L_y $).  For $J=0$ (broken line) where only the entropic repulsion exists between the steps, the $P(l)$ drops around $l \sim 0$ or $l \sim L_x $, with no peaks.
It is clear, therefore, that the attraction exists between steps for $J \neq 0$ case.

Origin of this attractive force is the fluctuation of the Ising spins (number density fluctuation of the adsorbates).
After coarse graining, Ising system can be described in terms of the $\phi^4$-like continuous ``massive'' field\cite{amit} except for at criticality (the transition temperature of the Ising system at zero field is $0.15 \times 2/\ln (1+\sqrt{2}) \sim 0.3404$).  
Due to the coupling term of the form $-\alpha \sigma |\Delta h|$, there appears effective force between the height differences $\{\Delta h\}$.
The force mediated by the massive field is short-ranged, with the range being $\sim \xi $ ($\xi $: correlation length of the Ising model).  
In other words, the adsorbates play the role of photons (for Coulomb force) or Yukawa mesons (for nuclear force). 
This attractive interaction leads to the instability against step bunching.

At the end of the first-order transition line, there usually appears a critical point where the second order transition occurs.  In Ref. \cite{aay99} We find that $B$  becomes zero at a temperature  where the surface free energy has the  form  
\beq
f(\rho)=f(0)+\gamma \rho + C \rho^5+(\mbox{higher order}).
\eeq
The GMPT behavior (\ref{GMPTform}) breaks down at this point.
In \cite{aay99}, we have presented two possible mechanisms for the vanishing of  $B$ (``stiffening transition''): (1) ``single-particle'' mechanism, and (2) intrinsically many-body mechanism.
From the present study, we conclude that the second mechanism is the case, and that the ``stiffening transition'' is naturally interpreted as the one which is very much like the critical point in the liquid-gas transition.

\acknowledgments

  This work was partially supported by the ``Research for the Future'' Program from The Japan Society for the Promotion of Science (JSPS-RFTF97P00201) and by the Grant-in-Aid for Scientific Research from Ministry of Education, Science, Sports and Culture (No.09640462).

\begin{figure}
\caption{
(a) The vertex of the RSOS model.    A vertex is drawn along the boundary between four adjacent SOS columns.  At the legs of the vertex, variables $\mu$, $\chi $, $\delta $ and $\nu $ are assigned as $h_2-h_1$, $h_1 - h_3$, $h_4 - h_3$ and $h_2 - h_4$, respectively. These variables take $\{1, 0, -1\}$ and should satisfy $\mu + \chi = \delta + \nu $.  
(b) The vertex representation of the Ising-RSOS coupled system.
Ising spins which take $\pm 1$ are described by open circles on the legs of vertex. 
}
\label{fig1}
\end{figure}

\begin{figure}
\caption{
The PWFRG results for the $p-\beta \eta$ curves with $\alpha=0.5$, $H/\epsilon = -0.01$ and $J=0.15$ where $\beta=1/\kBT$.
 The temperature of each curve is $\kBT/\epsilon = 0.3, 0.4$,  and $0.5$ from  right to  left.
}
\label{fig2}
\end{figure}

\begin{figure}[htbp]
\caption{
The snapshot of the Monte Carlo simulation at $10^6$ MCS.
Above: Configuration of RSOS vicinal surface; a black dot shows $\Delta h= -1$, a white dot shows $\Delta h= 1$, and gray region corresponds to $\Delta h= 0$.
Below: Configuration of Ising system (adsorption layer); a white dot shows $\sigma = 1$ (occupied)  and a black dot shows $\sigma = - 1$ (empty).
$\alpha=0.5$. $H/\epsilon = -0.01$.  $\kBT/\epsilon = 0.4$.
Size: $408 \times 120$. We have 24 steps on the surface.
}
\label{fig3}
\end{figure}

\begin{figure}
\caption{
The ``terrace width distribution'' calculated by the Monte Carlo method averaged over $4 \times 10^7$MCS. $\alpha=0.5$. $H/\epsilon = -0.01$.  $\kBT/\epsilon = 0.4$.
Size: $70 \times 120$. 
The bold line and the broken line shows $J=0.15$ and $J=0$, respectively.
}
\label{fig4}
\end{figure}

\end{document}